\documentstyle[aps,eqsecnum,psfig]{revtex}

\begin{document}
\draft
\title{Neutron capture of $^{26}$Mg at thermonuclear energies}
\author{P.~Mohr$^{1,2}$, H.~Beer$^3$, H.~Oberhummer$^2$, and G.~Staudt$^4$
}
\address{
        $^1$ Institut f\"ur Kernphysik, Technische Universit\"at Darmstadt,
        Schlossgartenstr.~9, D--64289 Darmstadt, Germany \\
        $^2$ Institut f\"ur Kernphysik, Technische Universit\"at Wien,
        Wiedner Hauptstr.~8-10, A--1040 Vienna, Austria \\
        $^3$Forschungszentrum Karlsruhe, Institut f\"ur Kernphysik III,
        P.~O.~Box 3640, D--76021 Karlsruhe, Germany \\
        $^4$Physikalisches Institut, Universit\"at T\"ubingen,
        Auf der Morgenstelle 14, D--72076 T\"ubingen, Germany
}

\date{\today}
\maketitle
\begin{abstract}
The neutron capture cross section of $^{26}$Mg was measured 
relative to the known gold cross section
at thermonuclear energies using the fast cyclic activation technique.
The experiment was performed at the 3.75 MV Van-de-Graaff accelerator,
Forschungszentrum Karlsruhe. 
The experimental capture cross section is the sum of resonant
and direct contributions. For the resonance at 
$E_{\rm{n,lab}} = 220~{\rm{keV}}$ our new results are in disagreement
with the data from Weigmann {\it et al}.
An improved Maxwellian averaged capture cross section is derived 
from the new experimental data taking into account $s$- and $p$-wave
capture and resonant contributions.
The properties of so-called potential resonances which influence the
$p$-wave neutron capture of $^{26}$Mg are discussed in detail.
\end{abstract}




%
\section{Introduction}
\label{sec:intro}
Neutron capture processes of neutron--rich light isotopes
play an important role in astrophysical scenarios ranging from the
so--called
inhomogeneous Big Bang models to nucleosynthesis in stellar 
helium and carbon burning stages. 
In Inhomogeneous Big Bang Models the
high neutron flux induced primordial nucleosynthesis bridges the mass 5 and 
mass 8 gap~\cite{app88}. Subsequent neutron capture
processes of neutron--rich isotopes may even
lead to a primordial r--process~\cite{cow91,rau94}.
The efficiency for the production of heavy elements in such a scenario 
depends sensitively on the respective neutron capture rates for these
light isotopes. Therefore the neutron capture cross sections have to be 
determined over a wide energy range up to about 1 MeV.
 
In massive Red Giant stars magnesium isotopes are mainly
the products of hydrostatic carbon and neon burning. $^{25}$Mg
and $^{26}$Mg have also appreciable abundances in the
outer carbon layer as a result of the reactions
$^{22}$Ne($\alpha$,$\gamma$)$^{26}$Mg, $^{22}$Ne($\alpha$,n)$^{25}$Mg, and
$^{25}$Mg(n,$\gamma$)$^{26}$Mg~\cite{woo95}.
He--shell burning in low mass asymptotic giant branch (AGB) stars
has been proposed as the site for the main component of the s--process
\cite{str95,gal97}.
In AGB stars $^{26}$Mg is likely to be made by
$^{22}$Ne($\alpha$,$\gamma$)$^{26}$Mg,
as well as by $^{25}$Mg(n,$\gamma$)$^{26}$Mg, by the decay
of $^{26}$Al, and by $^{26}$Al(n,p)$^{26}$Mg with radioactive
$^{26}$Al being synthesized through 
the previous H--burning shell. 
For the above discussed stellar scenarios
neutron capture rates need to be known in the energy range between
about 5 and 200 keV.

The reaction rate for $^{26}$Mg(n,$\gamma$)$^{27}$Mg
that is being investigated in this work is low through its small
cross section. The reaction $^{26}$Mg(n,$\gamma$)$^{27}$Mg is
of interest in stellar nucleosynthesis, because
(i) $^{26}$Mg is one of the most abundant isotopes in the cosmos,
(ii) all Mg--isotopes show a strong deviation from the normal 1/v--behavior
of their cross sections, and
(iii) of the interplay between neutron production and captures for
Mg--isotopes, and with the concurrential radioactive $^{26}$Al.

The cross section for neutron capture processes is dominated by
the nonresonant direct capture process (DC) and by contributions from
single resonances which correspond to neutron unbound states in the
compound nucleus (CN). For calculating the different reaction contributions
we used a simple hybrid model: the nonresonant contributions were
determined by using a direct capture model, the resonant contributions
were based on determining the resonant Breit Wigner cross section. In the
case of broad resonances additional interference terms have to be taken 
into account which were neglected in the present work because the relevant
resonances are relatively narrow.

In this work we also discuss the
effects of potential resonances in direct capture.
Potential resonances are well known in nuclear processes 
(see Ref.~\cite{ohu91} and references therein), 
but their influence
in numerical direct--capture calculations has to our knowledge not been
investigated before in great detail. 
The reaction $^{26}$Mg(n,$\gamma$)$^{27}$Mg
provides an excellent example for discussing the artifacts
such resonances can produce in direct--capture calculations.

Up to now the neutron capture cross section of $^{26}$Mg was measured 
at thermal energies in several experiments 
(see Refs.~\cite{wal92,sel70,ryv70}, and references therein).
However, at thermonuclear energies only the resonance properties 
of four resonances in $^{26}$Mg(n,$\gamma$)$^{27}$Mg were
determined by Weigmann {\it et al.} \cite{weig76}, and in that work
the Maxwellian averaged capture cross section 
$< \sigma > \, = \, < \sigma \cdot v > \, / \, v_{\rm T}$,
with $v_{\rm T} = \sqrt{2 k_B T / \mu}$, was calculated by the sum of the
$s$-wave contribution (extrapolated from the thermal capture cross
section assuming the usual $1/v$ law) and of the contributions from the
four measured resonances. The $p$-wave contribution of the DC was neglected
in that work.

In the following we will present our experimental procedure and results
in Sect.~\ref{sec:exp}. In Sect.~\ref{sec:theo} we analyze our new
experimental data using the DC model and we discuss the properties of
so-called potential resonances, 
in Sect.~\ref{sec:astro} we show the improved Maxwellian averaged
capture cross section $< \sigma >$,
and the results are discussed
in Sect.~\ref{sec:summ}.

\section{Experimental Setup and Results}
\label{sec:exp}
The experiment was performed at the 3.75 MV Van-de-Graaff accelerator
of the Forschungszentrum Karlsruhe. The residual nucleus
$^{27}$Mg decays with a half-life of $T_{1/2} = 9.458~{\rm{min}}$
to $^{27}$Al. Two $\gamma$ rays with
$E_{\gamma} = 843.76~{\rm{keV}}$ (branching 71.8 \%)
and 
$E_{\gamma} = 1014.44~{\rm{keV}}$ (branching 28.0 \%)
can be detected \cite{endt90}. The decay properties 
of $^{27}$Mg and $^{198}$Au (from the neutron capture of $^{197}$Au)
are listed in 
Table \ref{tab:tab1}.

\subsection{Fast cyclic activation technique}
\label{subsec:fast}
For residual nuclei with short half-lives the fast cyclic activation technique
was developed at the Forschungszentrum Karlsruhe \cite{beer94}.
In the $^{26}$Mg experiment
the sample is irradiated for a period $t_{\rm b}$, 
after this irradiation time
the sample is moved to the counting position in front of a
high--purity germanium (HPGe) 
detector (waiting time $t_{\rm{w_1}}$).
The $\gamma$--rays following the $\beta$--decay of $^{27}$Mg are
detected for a time interval $t_{\rm c}$, 
and finally the sample
is moved back to the irradiation position 
(waiting time $t_{\rm{w_2}}$).
The whole cycle with duration 
$T = t_{\rm b} + t_{\rm{w_1}} + t_{\rm c} + t_{\rm{w_2}}$
is repeated many times to gain statistics.
The $^{26}$Mg(n,$\gamma$)$^{27}$Mg experiment was performed partly with
$t_{\rm b} = 49.6~{\rm{s}}$, 
$t_{\rm{w_1}} = 0.4~{\rm{s}}$, 
$t_{\rm c} = 49.6~{\rm{s}}$, 
$t_{\rm{w_2}} = 0.4~{\rm{s}}$,
$T = 100~{\rm{s}}$,
partly with
$t_{\rm b} = 119.4~{\rm{s}}$, 
$t_{\rm{w_1}} = 0.6~{\rm{s}}$, 
$t_{\rm c} = 119.4~{\rm{s}}$, 
$t_{\rm{w_2}} = 0.6~{\rm{s}}$, 
$T = 240~{\rm{s}}$.

The accumulated number of counts from a total of $N$ cycles,
$C=\sum_{i=1}^n C_i$, where $C_i$,
the counts of the i-th cycle, are calculated for a chosen irradiation
time, $t_{\rm b}$, which is short
enough compared with the fluctuations of the neutron flux, 
is~\cite{beer94}
\begin{equation}
\label{eq:eq1}
C  = \epsilon_{\gamma}K_{\gamma}f_{\gamma}\lambda^{-1}[1-\exp(-\lambda
t_{\rm c})]
\exp(-\lambda t_{\rm{w_1}})
 \frac{1-\exp(-\lambda t_{\rm b})}{1-\exp(-\lambda T)} N \sigma_\gamma
{[1-f_{\rm b} \exp(-\lambda T)]}
 \sum_{i=1}^n \Phi_i
 \end{equation}
with
\begin{equation}
        f_{\rm b}  =  
                \frac{\sum_{i=1}^n \Phi_i \exp[-(n-i)\lambda T]}
                        {\sum_{i=1}^n \Phi_i} \quad .
\end{equation}
The following additional quantities have been defined:
$\epsilon_\gamma$: detection efficiency, $K_\gamma$:
$\gamma$-ray absorption, $f_\gamma$: $\gamma$-ray intensity per decay,
$\lambda$: decay constant,
$N$: the thickness (atoms per barn) of target
nuclei, $\sigma_\gamma$: the capture cross section, $\Phi_i$: the neutron flux
in the i-th cycle. The
quantity $f_{\rm b}$ is calculated from the 
registered flux history of a $^6$Li
glass monitor which is mounted at a distance of
91 cm from the neutron production target.
A simple n--$\gamma$ discrimination was applied to the slow output 
of the $^6$Li monitor: the amplitude of the neutron signal from
the $^6$Li(n,$\alpha$)$^3$H reaction is much larger than the amplitude
from a typical $\gamma$ ray event
that cannot deposit its full energy in the 1 mm thin
$^6$Li glass scintillator.

The experimental uncertainties are reduced to a great amount by measuring
the $^{26}$Mg capture cross section relative to the well-known gold
standard \cite{mack75}. For this purpose our magnesium samples were
sandwiched between two thin gold foils. Then the capture cross section
of the sample $\sigma_\gamma^{\rm S}$ is given relative to the
capture cross section of the gold reference $\sigma_\gamma^{\rm R}$ by:
\begin{equation}
\label{eq:eq2}
\sigma_\gamma^{\rm S} = \sigma_\gamma^{\rm R} \;
        \frac{
                C^{\rm S} N^{\rm R}
                \epsilon_{\gamma}^{\rm R}K_{\gamma}^{\rm R}f_{\gamma}^{\rm R}
                \lambda^{\rm S}[1-\exp(-\lambda^{\rm R}t_{\rm c})]
                \exp(-\lambda^{\rm R}t_{\rm{w_1}})
                [1-\exp(-\lambda^{\rm R}t_{\rm b})]
                [1-\exp(-\lambda^{\rm S}T)]
                {[1-f_{\rm b}^{\rm R}\exp(-\lambda^{\rm R}T)]}
                }
             {
                C^{\rm R} N^{\rm S}
                \epsilon_{\gamma}^{\rm S}K_{\gamma}^{\rm S}f_{\gamma}^{\rm S}
                \lambda^{\rm R}[1-\exp(-\lambda^{\rm S}t_{\rm c})]
                \exp(-\lambda^{\rm S}t_{\rm{w_1}})
                [1-\exp(-\lambda^{\rm S}t_{\rm b})]
                [1-\exp(-\lambda^{\rm R}T)]
                {[1-f_{\rm b}^{\rm S}\exp(-\lambda^{\rm S}T)]}
                }
\end{equation}

\subsection{Samples}
\label{subsec:samples}
Our samples consisted of MgO powder enriched in $^{26}$Mg by 98.79 \% and of
metallic magnesium of natural isotopic composition. 
The MgO powder was pressed to two small tablets with
a diameter of $d = 6~{\rm{mm}}$ and a thickness of about 0.5 mm and 1.5 mm,
respectively. 
The MgO tablets were put into a thin Al foil to ensure that no powder
is lost during the measurements. The mass of the tablets had to be
determined very carefully because the MgO powder is very hygroscopic.
Therefore, the tablets were heated at $T \approx 1000^\circ~{\rm{C}}$ for
several hours, and then they were weighted during the cooling phase.
We removed the tablets from the oven at about $300^\circ~{\rm{C}}$. 
After this procedure
we found that the measured weight of the tablets was stable for a few
minutes, but then the weight began to increase by about 10 to 20 percent
within a few hours.

As a further check we repeated one activation measurement with a sample
of metallic $^{\rm{nat}}$Mg with the same geometric shape (diameter 6 mm,
thickness 1 mm, mass 47.77 mg). The result agreed perfectly with the cross
section measured with the enriched samples. This confirms that the
weighting procedure worked well, that the enrichment
of $^{26}$Mg in the MgO samples was correct, and
that the influence of degraded neutrons scattered by hydrogen
in the hygroscopic MgO samples is negligible. 
The properties of the samples are
summarized in Table \ref{tab:tab2}.

\subsection{Neutron production, Time-of-Flight measurements}
\label{subsec:neutron}
The neutrons for the activation were generated using the $^7$Li(p,n)$^7$Be
reaction. We measured the capture cross section in the astrophysically
relevant energy region of several keV at six different neutron energies
which are obtained by different combinations of proton energy and 
thickness of the lithium target:
($i$) Close above the reaction threshold at 
$E_{\rm{p,lab}} = 1881~{\rm{keV}}$ one obtains a quasi--Maxwellian
spectrum with $k_B T = 25~{\rm{keV}}$ \cite{rat88} if the energy loss of the
protons in the lithium layer of the target reduces the proton energy below the 
reaction threshold. 
($ii$) Very close to the reaction threshold a narrow spectrum with
$12~{\rm keV} \le E_{\rm{n,lab}} \le 52~{\rm keV}$ was obtained.
Note that in these cases ($i$) and ($ii$) the neutron energy spectrum 
is independent 
of the thickness of the lithium layer. We used a lithium layer of 
30 $\mu$m which is by far thick enough for that purpose.
($iii$)
Using four different higher proton energies we obtained four different 
neutron energy spectra with higher neutron energies and an energy width 
of about 20--50 keV. Now the energy width
depends sensitively on the thickness of the lithium layer which was 
about 2.6 $\mu$m in our experiment. 
The thicknesses of the different thin lithium targets were determined by two
independent methods: first during the evaporation
of the metallic lithium layer and second from the width of the 
time-of-flight spectra (see below).
Both methods give the same results within an uncertainty of about 10\%.

For the activation runs beam currents in the order of
80 to 100 $\mu$A were used. Because of the electric power 
of the proton beam of about
$P \approx 2~{\rm{MV}} \cdot 100~{\rm{\mu A}} \approx 200~{\rm{W}}$
the copper backing of the thin lithium layer was water-cooled, and
the temperature of the copper backing remained below 85$^o$C during the
whole experiment.

\begin{figure}
\begin{center}
        \leavevmode
        \psfig{figure=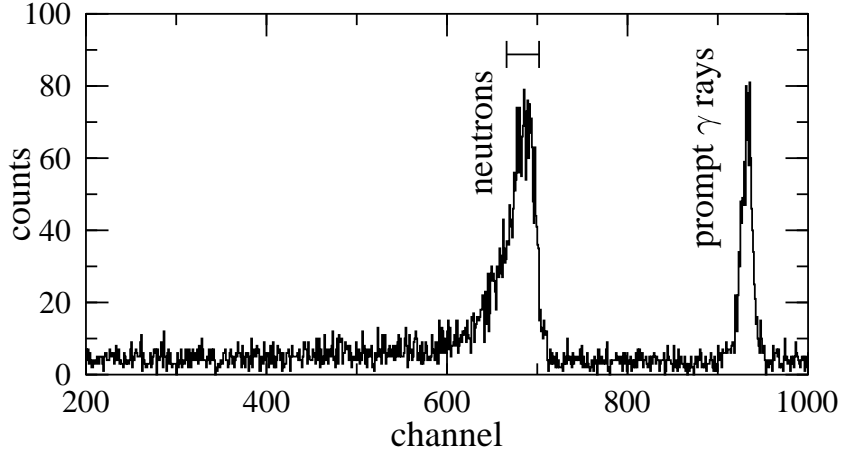,bbllx=110pt,bblly=100pt,bburx=500pt,bbury=360pt}
\end{center}
\caption{
	\label{fig:tof} 
	Typical neutron TOF spectrum from a thin lithium target
	(thickness 2600 nm $\pm$ 10\%)
	at $\vartheta = 0^\circ$ and at a
	distance of $D = 91.3~{\rm{cm}}$. One can see
	the narrow peak from prompt $\gamma$ rays at channel 932.
	The broad neutron peak starts at channel 702 
	($E_{\rm{n,lab}}^{\rm max} = 134.5~{\rm{keV}}$),
	and the tail of the neutron energy distribution ends
	at channel 615
	($E_{\rm{n,lab}}^{\rm min} = 71.5~{\rm{keV}}$).
	The bar over the neutron peak indicates the expected energy
	width calculated from the energy loss of the protons
	\protect\cite{ziegler} and from the reaction kinematics
	neglecting the low-energy tail from the diffusion of lithium
	into the copper backing.
	The time resolution is about 10 ns (FWHM of the $\gamma$ peak).
}
\end{figure}

The neutron energy spectra were controlled by time-of-flight (TOF)
measurements using the pulsed beam which is available from the Van-de-Graaff
accelerator. Typically the repetition frequency is 1 MHz, and the pulse width
is about 10 ns. The beam current in the pulsed mode was about 1 to 4 $\mu$A.
The neutrons were detected in the $^6$Li glass monitor, and the TOF of the
neutrons was measured between the fast timing output from the $^6$Li monitor
(start) and the delayed pickup signal from the pulsed beam (stop). 
Especially at energies close to the resonances in the 
$^{26}$Mg(n,$\gamma$)$^{27}$Mg reaction the energy of the neutrons has to be
determined very accurately. A typical TOF spectrum at 
$\vartheta_{\rm{lab}} = 0^\circ$
is shown in Fig.~\ref{fig:tof}.

The result of our experiment is an average capture cross section
\begin{equation}
\bar{\sigma} 
  = \int \sigma(E_{\rm{n,lab}}) \Phi(E_{\rm{n,lab}}) dE_{\rm{n,lab}}
\label{eq:sigma_bar}
\end{equation}
where the neutron energy distribution $\Phi(E_{\rm{n,lab}})$ 
is normalized to 1:
\begin{equation}
\int \Phi(E_{\rm{n,lab}}) dE_{\rm{n,lab}} = 1
\label{eq:phi}
\end{equation}

The neutron energy distribution $\Phi(E_{\rm{n,lab}})$ was calculated 
from the measured TOF distribution
$\Phi(T)$ by the relation $\Phi(T) dT = \Phi(E_{\rm{n,lab}}) dE_{\rm{n,lab}}$.
Additionally several facts have to be taken into account: 
($i$) the angular dependence of the $^7$Li(p,n)$^7$Be reaction cross section
which was measured at several angles 
using the $^6$Li monitor \cite{mohr_oakridge},
($ii$) the geometry of the neutron production target and the sample,
($iii$) the energy loss of the protons in the lithium layer with a
thickness $d = 2600~{\rm{nm}} \pm 10\%$ \cite{ziegler},
($iv$) the low-energy tail of the neutron energies in the TOF spectrum
(see Fig.~\ref{fig:tof}). ($iii$) and ($iv$) are relevant only for
the measurements at higher neutron energies.
The neutron energy distribution with 
$\bar{E}_{\rm{n,lab}} = 208.3~{\rm{keV}}$
is shown in Fig.~\ref{fig:phi}.
It has to be pointed out that it is difficult to measure the effective
neutron energy distribution $\Phi(E_{\rm{n,lab}})$ directly
because of the extended
lithium target and the extended sample. However, all ingredients 
($i$)--($iv$)
for the 
calculation of $\Phi(E_{\rm{n,lab}})$ were determined experimentally, and
so the uncertainties of the calculation are small.

\subsection{$\gamma$ ray detection}
\label{subsec:gamma}
\begin{figure}
\begin{center}
        \leavevmode
        \psfig{figure=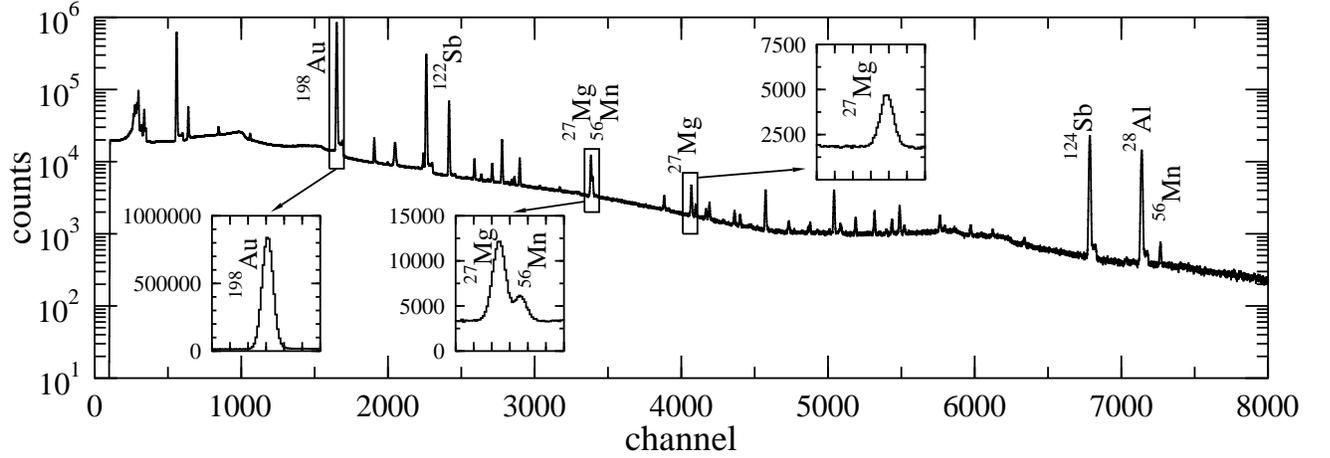,bbllx=200pt,bblly=100pt,bburx=400pt,bbury=300pt}
\end{center}
\caption{
	\label{fig:spec} 
	Energy spectrum of the HPGe detector measured during
	the activation of $^{26}$Mg.
	In the insets the
	relevant areas around 411.80 keV (decay of $^{198}$Au),
	and 843.76 and 1014.44 keV (decay of $^{27}$Mg) are shown.
	The main background lines come from the decay of
	$^{56}$Mn 
	($E_\gamma = 846.76~{\rm{keV}}$, only 3 keV above the 
	$^{27}$Mg decay line), from $^{28}$Al,  and from $^{122,124}$Sb.
	Note the logarithmic scale of the full spectrum and the
	linear scale of the three insets. 
}
\end{figure}

The $\gamma$ rays from the decay of $^{27}$Mg and $^{198}$Au were
detected using a high purity germanium (HPGe) detector with a
relative efficiency of 60 \% (compared to a 3'' x 3'' NaI detector).
The energy-dependent efficiency of the HPGe detector
was calculated using the detector simulation program GEANT \cite{GEANT}.
The simulation calculation was controlled by an experimental determination
of the efficiency with several radioactive calibration sources.
The results for the relative efficiencies are
\begin{eqnarray}
\frac{
        \epsilon_\gamma (411.80~{\rm{keV}})
        }
     {
        \epsilon_\gamma (843.76~{\rm{keV}})
        } = 1.752 \pm 2 \% & \,\,\,\, \mbox{and} \,\,\,\, &
\frac{
        \epsilon_\gamma (411.80~{\rm{keV}})
        }
     {
        \epsilon_\gamma (1014.44~{\rm{keV}})
        } = 2.016 \pm 2 \% \,\,\,\mbox{for}\,\,\,d = 30~{\rm{mm}}\nonumber \\
\frac{
        \epsilon_\gamma (411.80~{\rm{keV}})
        }
     {
        \epsilon_\gamma (843.76~{\rm{keV}})
        } = 1.746 \pm 2 \% & \,\,\,\, \mbox{and} \,\,\,\, &
\frac{
        \epsilon_\gamma (411.80~{\rm{keV}})
        }
     {
        \epsilon_\gamma (1014.44~{\rm{keV}})
        } = 2.007 \pm 2 \% \,\,\,\mbox{for}\,\,\,d = 35~{\rm{mm}}\nonumber
%
\end{eqnarray}
for the close geometry used in our experiment (the distance between sample
and detector was 30 mm for the $k_B T = 25~{\rm{keV}}$ runs and for the
run very close above the threshold, and 35 mm for the other runs).
Additionally,
the $\gamma$--ray absorption in the sample and in the gold foils
was calculated using the cross section tables provided by 
National Nuclear Data Center,
Brookhaven National Laboratory, via WWW, and based on Ref.~\cite{cullen}.
This small correction is in the order of a few \%.
A typical spectrum from the HPGe detector is shown in Fig.~\ref{fig:spec}.

\subsection{Experimental Results}
\label{subsec:results}
\begin{figure}
\begin{center}
        \leavevmode
        \psfig{figure=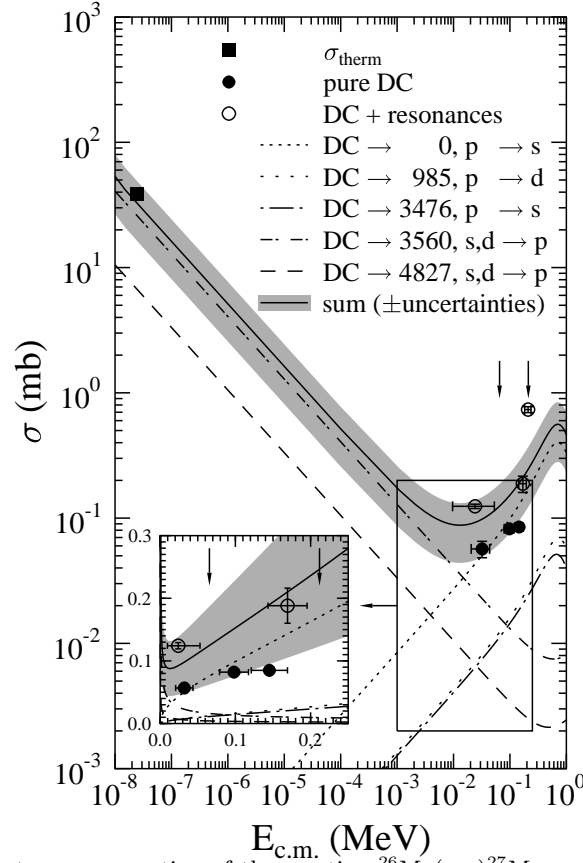,bbllx=50pt,bblly=120pt,bburx=400pt,bbury=420pt}
\end{center}
\caption{
	\label{fig:exp} 
	Experimental capture cross section of the reaction
	$^{26}$Mg(n,$\gamma$)$^{27}$Mg compared to a DC calculation.
	All circles are the result of this experiment.
	The full circles ($\bullet$) show the non-resonant part, 
	the open circles
	($\circ$) are influenced by the known resonances at
	$E_{\rm{n,lab}} = 68.7~{\rm{keV}}$
	and
	$E_{\rm{n,lab}} = 220~{\rm{keV}}$
	\protect\cite{weig76} which are indicated by the vertical arrows.
	The thermal capture cross section was taken from 
	Ref.~\protect\cite{wal92}.
	The horizontal error bars show the FWHM of the
	neutron energy spectra.
	Note the logarithmic scale of the big figure and the
	linear scale of the inset; the data point at
	$E_{\rm{n,lab}} = 208.3~{\rm{keV}}$ does not fit
	in the linear inset.
}
\end{figure}

The new experimental data are listed in Table \ref{tab:tab3}, and they are
compared to a DC calculation (see Sect.~\ref{sec:theo}) in 
Fig.~\ref{fig:exp}. Two of the four known resonances at 
$E_{\rm{n,lab}} = 68.7~{\rm{keV}}$, 220 keV, 427 keV, and 432 keV \cite{weig76}
have influence on our measured cross sections at $k_B T = 25~{\rm{keV}}$ 
(resonance at 68.7 keV) and at
$E_{\rm n,lab} = 178~{\rm{keV}}$ and $208~{\rm{keV}}$
(resonance at 220 keV). The two resonances above 400 keV do not
influence our measurement because they are narrow ($\Gamma \le 2~{\rm{keV}}$)
\cite{weig76}.

The influence of a narrow resonance at the energy $E_{\rm{res}}$
on the measured capture cross section can be calculated by
\begin{equation}
\bar{\sigma}_{\rm{res}} 
  = \int \sigma_{\rm{BW}}(E_{\rm{c.m.}}) \Phi(E_{\rm{c.m.}}) dE_{\rm{c.m.}}
  \approx \Phi(E^{\rm{res}}_{\rm{n,lab}}) \;
        \frac{A_P + A_T}{A_T} \; 2\pi \; 
        \frac{\pi}{{(k^{\rm{res}}_{\rm{c.m.}})}^2} \; (\omega \gamma)
\label{eq:sigma_BW}
\end{equation}
where 
$\sigma_{\rm{BW}}(E_{\rm{c.m.}})$ is the Breit-Wigner shaped
capture cross section in the resonance,
$\Phi(E)$ the neutron energy distribution,
$k^{\rm{res}}_{\rm{c.m.}}$ the wave number at the resonance energy 
$E_{\rm{res}} = \hbar^2 (k^{\rm{res}}_{\rm{c.m.}})^2 / 2 \mu$
in the center-of-mass system,
and $\omega \gamma$ is the resonance strength.

\begin{figure}
\begin{center}
        \leavevmode
        \psfig{figure=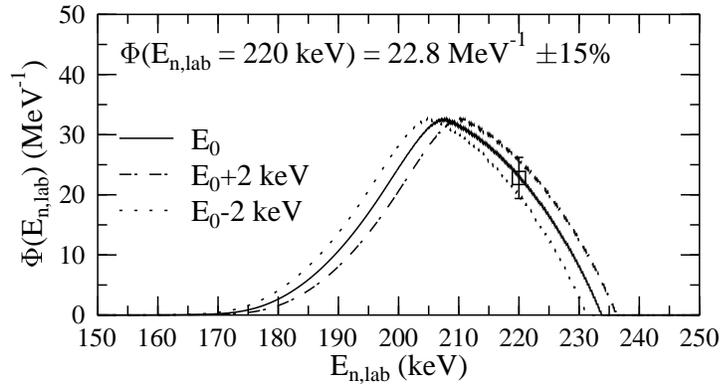,bbllx=110pt,bblly=100pt,bburx=400pt,bbury=300pt}
\end{center}
\caption{
	\label{fig:phi} 
	Calculated neutron energy distribution $\Phi(E_{\rm{n,lab}})$ with
	$\bar{E_{\rm n}} = 208.3~{\rm{keV}}$ which was used for the
	measurement of the $E_{\rm{n,lab}} = 220~{\rm{keV}}$
	resonance \protect\cite{weig76}. 
	For comparison we also show the energy distributions which
	are obtained from a by 2 keV increased/decreased primary proton energy
	(dashed lines).
	The value at the resonance energy is 
	$\Phi(E_{\rm{n,lab}} = 220~{\rm{keV}}) = 22.8~{\rm{MeV^{-1}}}
	\pm 15\%$ (indicated by the data point).
}
\end{figure}

The resonance strength of the 
$E_{\rm{n,lab}} = 220~{\rm{keV}}$
resonance was measured by Weigmann {\it et al.} \cite{weig76}:
$\omega \gamma = 3.75 \pm 0.16~{\rm{eV}}$. 
This resonance gave the dominating contribution
to the capture cross section for the neutron energy spectrum
$\Phi(E_{\rm{n,lab}})$ 
shown in Fig.~\ref{fig:phi}
(corresponding to $E^0_{\rm{p,lab}} = 2003~{\rm{keV}}$).
>From Eq.~\ref{eq:sigma_BW} one obtains
$\bar{\sigma}_{\rm{res}} = 1.78 \pm 0.27~{\rm{mb}}$.
In contradiction to that result our measured capture cross section
is only $\sigma = 0.735 \pm 0.040~{\rm{mb}}$ as shown in Fig.~\ref{fig:exp}!

Our method for measuring neutron capture cross sections is mainly tailored
for non-resonant capture. Nevertheless, our experimental result definitely
excludes the result of Weigmann {\it et al.} \cite{weig76} for the resonance
at 220 keV.
To estimate the uncertainties of our calculated $\Phi(E_{\rm{n,lab}})$
we calculated energy distributions $\Phi(E_{\rm{n,lab}})$ 
also at proton energies
$E_{\rm{p,lab}} = E^0_{\rm{p,lab}} \pm 2~{\rm{keV}}$ which is much larger
than the uncertainty of the accelerator calibration which was confirmed by
the TOF measurements. These results for 
$E_{\rm{p,lab}} = E^0_{\rm{p,lab}} \pm 2~{\rm{keV}}$ 
are also shown in Fig.~\ref{fig:phi} as dashed lines. 
The resulting uncertainties for $\Phi(E_{\rm lab}=220~{\rm{keV}})$ 
and $\bar{\sigma}_{\rm{res}}$ are about 15\%. 

If one subtracts
the expected non-resonant DC contribution 
$\sigma_{\rm DC} = 0.095 \pm 0.010~{\rm{mb}}$
(see also Sects.~\ref{subsec:DC} and \ref{sec:astro})
from our measured capture cross section at 
$E_{\rm{n,lab}} = 208.3~{\rm{keV}}$
($E_{\rm{c.m.}} = 200.3~{\rm{keV}}$)
we obtain a resonant contribution of $\bar{\sigma} = 0.64 \pm 0.05~{\rm{mb}}$,
which is a factor of about 2.8 lower than the result from 
Weigmann {\it et al.} \cite{weig76}. 
Therefore, we obtain a resonance strength of 
$\omega \gamma = 1.34 \pm 0.24~{\rm{eV}}$ for the
$E_{\rm{n,lab}} = 220~{\rm{keV}}$ resonance.
Interference effects between the DC and the
$E_{\rm{n,lab}}=220~{\rm{keV}}$ resonance
can be neglected because the total width of this resonance
is relatively small: $\Gamma = 0.44 \pm 0.07~{\rm{keV}}$ \cite{weig76}.

A similar analysis of the 68.7 keV resonance has relatively large
uncertainties because only the tail of the neutron energy spectrum 
with $k_B T = 25~{\rm{keV}}$ overlaps with this 
relatively weak resonance 
($\omega \gamma = 73 \pm 20~{\rm{meV}}$ from Ref.~\cite{weig76},
at least more than one order of magnitude smaller than the 
$E_{\rm{n,lab}} = 220~{\rm{keV}}$ resonance), 
and the
dominating contributions to this data point at $k_B T = 25~{\rm{keV}}$
come from $s$- and $p$-wave capture.

\section{Theoretical Analysis}
\label{sec:theo}
\subsection{Direct capture model and folding potentials}
\label{subsec:DC}
The theoretical analysis was performed within the Direct Capture (DC) model.
The DC cross section is given by \cite{kim87,mohr93}
\begin{equation}
\sigma_{\gamma i}^{{\rm DC}}
   = 
   \int {\rm d}\Omega\,2\,\left( \frac{e^2}
  {\hbar\,{\rm c}} \right) \left( \frac{\mu {\rm c}^2}{\hbar\,{\rm c}} \right)
  \left( \frac{k_\gamma}{k_a} \right)^3 \frac{1}{2\,I_A + 1}\,
  \frac{1}{2\,S_a + 1} \sum_{M_A\,M_a\,M_B,\,\pi_\gamma}
  \mid T_{M_A\,M_a\,M_B,\,\pi_\gamma} \mid^2 \quad.
\label{eq:DC}
\end{equation}
The quantities $I_A$, $I_B$ and $S_a$ ($M_A, M_B$ and $M_a$)
are the spins (magnetic quantum numbers) of the target nucleus $A$,
residual nucleus $B$ and projectile $a$, respectively.
The reduced mass in the entrance channel is given by $\mu$. The
polarization $\pi_\gamma$ of the electromagnetic radiation can be $\pm 1$. 
The wave
number in the entrance channel and for the emitted radiation is given
by $k_a$ and $k_\gamma$, respectively.
The transition matrices $T = T^{\rm E1} + T^{\rm E2} + T^{\rm M1}$ depend on
the overlap integrals
\begin{equation}
I^{{\rm E}{\cal L}/{\rm M}{\cal L}}_{l_b\,j_b\,I_B;l_a\,j_a} =
   \int\,{\rm d}r\,u_{l_b\,j_b\,I_B}(r)
   {\cal O}^{{\rm E}{\cal L}/{\rm M}{\cal L}}(r)\,\chi_{l_a\,j_a}(r) \quad
\label{eq:ac}
\end{equation}
The radial part of the bound state wave function in the exit channel and the
scattering
wave function in the entrance channel is given by $u_{l_b\,j_b\,I_B}(r)$ and
$\chi_{l_a\,j_a}(r)$, respectively. The radial parts of the electromagnetic
multipole operators are well-known. The calculation of
the DC cross sections has been performed using the code TEDCA~\cite{TEDCA}.

For the calculation of the scattering wave function $\chi(r)$ and the
bound state wave function $u(r)$ we used systematic folding potentials:
\begin{equation}
V(R) =
  \lambda\,V_{\rm F}(R)
  =
  \lambda\,\int\int \rho_a({\bf r}_1)\rho_A({\bf r}_2)\,
  v_{\rm eff}\,(E,\rho_a,\rho_A,s)\,{\rm d}{\bf r}_1{\rm d}{\bf r}_2
  \label{ad}
\end{equation}
with $\lambda$ being a potential strength parameter close
to unity, and $s = |{\bf R} + {\bf r}_2 - {\bf r}_1|$,
where $R$ is the separation of the centers of mass of the
projectile and the target nucleus.
The density of $^{26}$Mg has been derived from the measured
charge distribution \cite{vri87}, and the effective nucleon-nucleon
interaction $v_{\rm eff}$
has been taken in the DDM3Y parametrization \cite{kobos84}.
The resulting folding potential has a volume integral per interacting
nucleon pair $J_R = 500.41\,{\rm{MeV\,fm^3}}$ ($\lambda = 1$)
and an rms-radius $r_{\rm rms} = 3.676\,{\rm{fm}}$.
Details about the folding procedure can be found for instance in
\cite{abele93}, the folding potential has been calculated by using the
code DFOLD \cite{DFOLD}. The imaginary part of the potential
is very small because of the small flux into reaction channels
and can be neglected in our case. This fact will be discussed
in Sect.~\ref{subsec:imag}.
The potential strength parameter $\lambda$ can be adjusted either
to the binding energy $E_B$ (bound state wave function) or to the
scattering length ($s$-wave scattering at thermal energies).

The Pauli principle was taken into account for the bound state
wave functions by the condition
\begin{equation}
Q = 2N + L
\label{eq:Q}
\end{equation}
where $Q$, $N$, and $L$ are the number of oscillator quanta,
nodes in the wave function, and angular momentum number.
For even-parity states one has to take $Q = 2$ in the $sd$ shell,
and for odd-parity states $Q = 3$. The parameters $\lambda$ for several
bound state wave functions are listed in Table \ref{tab:spec}.
Note that the $\lambda$ values do not differ more than 30\% from unity 
for all bound states ($\lambda \approx 0.9$ for $Q = 2$ states with
positive parity,
$\lambda \approx 1.25$ for $Q = 3$ states with negative parity).

The total capture cross section is given by the sum over all
DC cross sections for each final state $i$,
multiplied by the spectroscopic factor $C^2 S$
which is a measure for the probability of finding
$^{27}$Mg in a ($^{26}$Mg $\otimes$ n) single-particle configuration
\begin{equation}
\sigma_\gamma = \sum_i C^2_i S_i \, \sigma_{\gamma i}^{\rm{DC}} \quad .
\label{eq:C2S}
\end{equation}

For the $^{26}$Mg(n,$\gamma$)$^{27}$Mg reaction one has to take into account
all final states below the neutron threshold of $^{27}$Mg.
However, in practice only E1 transitions to final states with large
spectroscopic factors are relevant in the keV energy region.
Additionally, at thermal energies M1 and E2 transitions contribute to
about 17\% to the thermal capture cross section \cite{wal92,sel70}.

\subsection{Spectroscopic factors}
\label{subsec:SF}
For the theoretical prediction of the neutron capture cross section
the knowledge of the spectroscopic factors (SF) is essential. These SF
have been determined from the DWBA analysis of different transfer
reactions like e.g.~$^{26}$Mg(d,p)$^{27}$Mg.
Unfortunately, the experimental SF from the different experiments only
agree within a factor of 2 (although these experiments claim uncertainties
in the order of 10--20\%). Additionally, SF from theoretical 
shell model calculations
are in average roughly a factor of 2 smaller than the experimental SF.
These discrepancies are listed in Table \ref{tab:spec}.

To reduce these uncertainties the following procedure was applied.
At thermal energies SF can be determined with high accuracy from the
ratio of the experimental capture cross section to the calculated DC cross
section. The DC cross section can be calculated with small uncertainties
because the uncertainties in the calculation of the scattering wave function
can be minimized by adjusting the potential strength (parameter $\lambda$)
to the neutron scattering length \cite{beer96}.
In the case of $^{26}$Mg 
the scattering lengths 
$b_{\rm bound} = 4.89 \pm 0.15~{\rm{fm}}$ \cite{mugh}
and
$b_{\rm free} = (b_{\rm bound} - Z \; b_{\rm{ne}}) \; A / (A+1) 
        = 4.725 \pm 0.15~{\rm{fm}}$
lead to $\lambda = 0.9694 \pm  0.024$.
Using this value for $\lambda$ we obtain
for the two dominating E1 transitions at thermal energies:
$C^2 S (3/2^-, 3560~{\rm{keV}}) = 0.360 \pm 0.054$ and
$C^2 S (1/2^-, 4827~{\rm{keV}}) = 0.260 \pm 0.039$\footnote{The 
numerical values
are slightly different from a previous work \cite{mohr_bene} because of
the improved experimental data \cite{wal92} compared to \cite{sel70}
and because of an
improved adjustment of the potential strength to the free coherent
scattering length $b_{\rm free}$ instead of the bound $b_{\rm bound}$.}.
The calculation includes the uncertainties of the potential strength
parameter $\lambda$, of the thermal capture cross section
($\sigma_{\rm{thermal}} = 39.0 \pm 0.8~{\rm{mb}}$ \cite{wal92}),
and of the branching of the thermal capture \cite{wal92,sel70}.

The result for the SF is in average 14.1\% smaller than the result of 
Meurders {\it et al.} \cite{meur74} obtained from (d,p) measurements
for the two odd-parity levels
for which the SF can be determined accurately as described above.
Therefore, we renormalized the experimental SF from Ref.~\cite{meur74}
by the factor $0.852 \pm 0.091$. 
The experimental data of Ref.~\cite{meur74} were used
because only in that experiment all relevant levels were analyzed.
The result for the DC cross section is compared to the
experimental results in
Fig.~\ref{fig:exp}.
One can clearly see the transition from the $1/v$-behavior ($s$-wave capture)
to the $v$-behavior ($p$-wave capture) at $E \approx 20~{\rm{keV}}$.
This transition is typical especially for neutron-rich nuclei in the 
$sd$ shell with many bound states with even parity and large spectroscopic
factors.

The overall agreement between the experimental data points which are
not affected by the known resonances and the DC calculation is quite good. 
However, our DC calculation overestimates the $p$-wave capture.
There are two possible explanations: (i) the spectroscopic factors
of the final states with even parity (which cannot be determined by the
procedure described above) are too large, (ii) the optical potential
for the $p$-wave differs slightly from the $s$-wave potential. 

\subsection{Potential resonances}
\label{subsec:pot}
In the following our discussion is restricted to folding potentials with
one free parameter $\lambda$. The same arguments also hold for 
Saxon-Woods or other potentials, but in many cases the discussion may be
more complicated because the larger number of parameters of the 
Saxon-Woods potential.

\begin{figure}
\begin{center}
        \leavevmode
        \psfig{figure=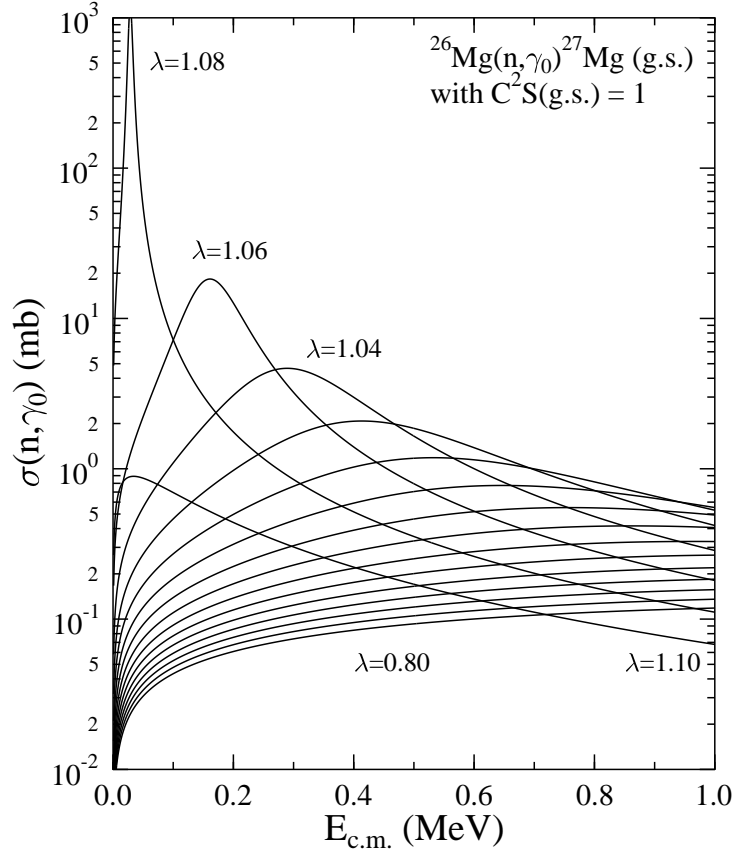,bbllx=50pt,bblly=90pt,bburx=400pt,bbury=450pt}
\end{center}
\caption{
	\label{fig:lambda} 
	Calculated capture cross section of the ground state transition
	$^{26}$Mg(n,$\gamma$)$^{27}$Mg$_{\rm g.s.}$
	with $C^2 S({\rm g.s.}) = 1.0$
	(E1 transition from the incoming $p$-wave to the bound $s$-wave).
	The capture cross section depends sensitively 
	on the potential strength
	parameter $\lambda$ of the incoming $p$-wave
	which was varied from $\lambda = 0.80$ (lowest
	capture cross section) up to $\lambda = 1.10$ in steps of 0.02
	(from bottom to top, exceptions are indicated in the figure).
	Note that the capture cross section varies up to five orders of
	magnitude when the potential strength is increased/decreased by
	about 15\% from the value for the $s$-wave
	($\lambda = 0.9694$)!
}
\end{figure}

The real part of the optical potential for the incoming
$p$-wave was assumed to be identical to the
$s$-wave potential which was adjusted to the neutron scattering length at
thermal energies. However, a weak
dependence (in the order of about 10--20\%)
of the optical potential on the parity and/or the angular momentum
of the partial wave was found in many systems 
(see e.g.~Ref.~\cite{mackintosh}).
This weak energy dependence has important consequences for the calculation
of the neutron capture cross section of $^{26}$Mg,
because the $^{26}$Mg-n potential shows a so-called
potential resonance for the $p$-wave
close to the astrophysically relevant energy region.
This fact leads to a very sensitive dependence of the capture cross section on
the potential strength (i.e.~the strength parameter $\lambda$).
In Fig.~\ref{fig:lambda} we show the dependence of the capture cross section
on the potential strength parameter $\lambda$. Changing $\lambda$ 
from 0.8 to about 1.0 does not change the qualitative shape 
of the cross section
curve, but the absolute value changes by a factor of 4.5 at 1 MeV. For
$\lambda > 1.0$ the resonant behavior becomes more and more visible,
and for $\lambda = 1.10$ the resonance disappears below the threshold
and leads to a bound state at $E_B = - 135.3~{\rm{keV}}$.

\begin{figure}
\begin{center}
        \leavevmode
        \psfig{figure=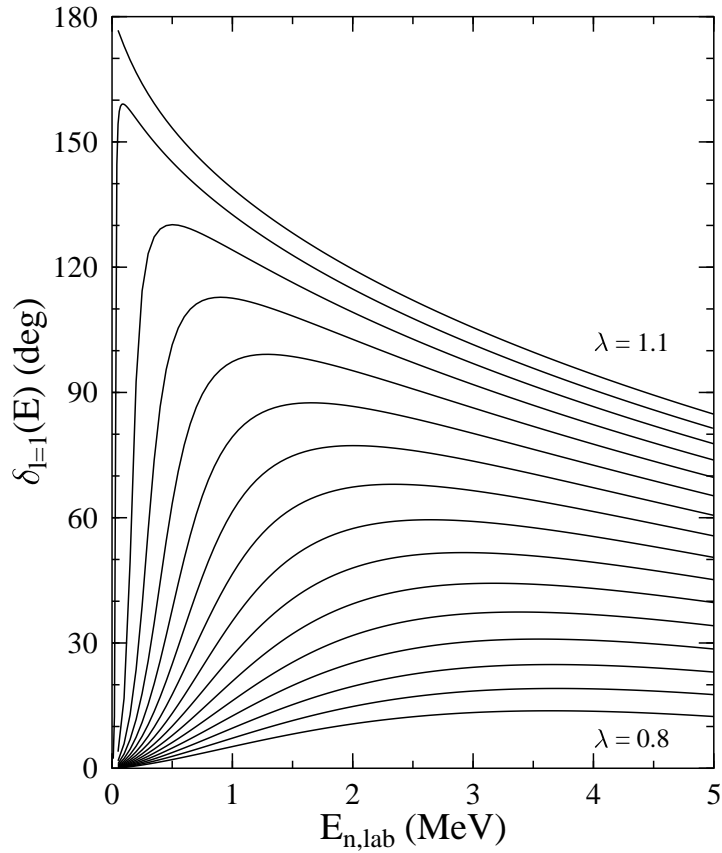,bbllx=50pt,bblly=100pt,bburx=400pt,bbury=450pt}
\end{center}
\caption{
	\label{fig:phase} 
	Scattering phase shifts $\delta_{l=1}(E)$ calculated for
	different potential strength parameters $\lambda$ from 0.80 to 1.10
	in steps of 0.02 (from bottom to top). 
	Note that the resonance position is shifted towards
	lower energies as the potential strength increases. 
	For $\lambda = 1.10$ the resonance disappears below the threshold
	(see also Table \ref{tab:res}).
}
\end{figure}

Such potential resonances are characterized by a large width because the simple
two-body potential model assumes full single-particle strength for
such resonances. For very broad resonances ($\Gamma \approx E$)
it is difficult to give precise numbers for the position and the
width of a resonance (see e.g.~the discussion in Ref.~\cite{barker}). 
To avoid such problems of definition we show the p-wave phase shift 
$\delta_{l=1}(E)$ for
several parameters $\lambda$ in Fig.~\ref{fig:phase}, and we define the
energy of the resonance as the maximum slope of the phase shift, and
the width is given by the slope at that energy:
$\Gamma = 2 \; [(d\delta_{l=1}(E)/dE)|_{E=E_{\rm{res}}}]^{-1}$.
The resulting values for $E_{\rm{res}}$, $\Gamma$, and the phase
shift $\delta_{l=1}(E_{\rm{res}})$ are listed in Table \ref{tab:res}.

For a proper adjustment of the $p$-wave potential strength the knowledge
of the experimental scattering phase shifts is desirable. However,
to our knowledge these phase shifts have not been measured.
Nevertheless, there is an experimental hint that such a potential resonance
is not only an artifact of the theory, but exists in nature: a broad
structure was observed in a measurement of the total $^{26}$Mg--n 
cross section at about 
$E \approx 300~{\rm{keV}}$ \cite{mugh}. At that energy our potential model
predicts a width of several hundred keV.

In general, several potential resonances have been proved experimentally.
A typical example is the first $3^+$ state in $^6$Li which can be found
as a resonance in the $^2$H($\alpha$,$\gamma$)$^6$Li capture 
reaction \cite{mohr94}. Because of the almost pure single-particle structure 
of $^6$Li = $^2$H $\otimes$ $\alpha$ the properties of this resonance
can be predicted by the potential model. For heavier nuclei an almost
pure single-particle structure cannot be expected but there still exist
broad resonances with a relatively strong single-particle component.
This leads to an overestimation of the width because the simple two-body
potential model assumes full single-particle structure for potential
resonances. As a consequence, the potential model is not able any more
to predict
the properties of such resonances with high accuracy; only if the
spectroscopic factor of such a resonance is known, the width can be
calculated using the relation
\begin{equation}
{\Gamma_{\rm{exp}}} = C^2 S \, {\Gamma_{\rm{s.p.}}}
\label{eq:SF}
\end{equation}
where $\Gamma_{\rm{s.p.}}$ is the width calculated from the potential model.
Furthermore,
the strength of the theoretical single-particle resonance is fragmented
into several states which can be identified experimentally. This can already
be seen from the fact that the two bound $p$-wave states which dominate
the thermal capture cross section of $^{26}$Mg(n,$\gamma$)$^{27}$Mg
have SF in the order of about 0.2 to 0.4.

It has to be pointed out that the so-called ``non-resonant'' energy region
can be calculated by the simple potential model {\bf only} if the potential
is carefully adjusted because the potential model {\bf always} contains
potential resonances, and as one can see from 
Figs.~\ref{fig:lambda} and \ref{fig:phase}, there
is no clear definition for ``resonant'' or ``non-resonant''.
A wrong adjustment may lead to ($i$) an overestimation
of the capture cross section if the potential contains a
single-particle resonance which does not exist in the experiment, and
($ii$) an underestimation of the capture cross section if the adjustment
neglects strong (and broad) single-particle resonances. 

The success of many previous calculations using the DC model, which seems
to be in contradiction to the above statements, can be explained by two
facts: ($i$) in several cases the influence of potential resonances is small
in the energy region which was analyzed in the calculation (this is
especially the case when the main contribution for the DC transition
comes from regions far outside the nucleus, 
see e.g.~Refs.~\cite{morlock97} or \cite{mengoni}),
($ii$) even wrong DC calculations can be corrected using 
adjustable spectroscopic
factors (uncertainties in the order of a factor of 2 are realistic,
see Sect.~\ref{subsec:SF}!).

>From this point of view the rough agreement of the calculated DC cross section
with the experimental results is already somewhat surprising because the
range of the capture cross section (see Fig.~\ref{fig:lambda}) covers
five orders of magnitude when the potential strength changes by only
$\pm 15\%$ from about 0.95 (determined from the $s$-wave scattering)
down to 0.8 and up to 1.1. Such a dramatic behavior of the so-called
``non-resonant'' direct capture can appear in any DC calculation.
To the best of our knowledge, these facts were not completely
discussed in all the
previous papers dealing with DC. In conclusion, we point out that the
calculation of the DC has to be done with great care for each nucleus
because otherwise deviations
between the predicted and the experimental
capture cross section  not only by a factor of two, but even
by orders of magnitude can appear, and both overestimations and
underestimations of the experimental cross sections are possible!

\subsection{The imaginary part of the optical potential}
\label{subsec:imag}
As pointed out by Lane and Mughabghab \cite{lane74},
the capture amplitude has to be derived from the full 
optical potential, i.e.\ real and imaginary part (see also
Eq.~\ref{eq:ac}). However, in the case of 
$^{26}$Mg(n,$\gamma$)$^{27}$Mg
the imaginary part of the optical potential can be neglected 
for the following reasons. The only open inelastic channel is 
neutron capture. Then the  capture cross section $\sigma_\gamma$
is directly related to the reflexion coefficient $\eta_l$ by:
\begin{equation}
\sigma_\gamma = 
  \frac{\pi}{{(k_{\rm{c.m.}})}^2}
  \sum_l (2 l + 1) ( 1 - \eta_l^2)~.
\label{eq:imag1}
\end{equation}
Using the real folding potential with
$\lambda = 0.9694$ (from Sect.~\ref{subsec:SF}),
an imaginary Saxon-Woods potential with depth $W_0$, 
reduced radius $R_0 = 1.3~{\rm{fm}}$
($R = R_0 \cdot A_T^{1/3}$),
diffuseness $a = 0.65~{\rm{fm}}$,
and the experimentally determined values of $\sigma_\gamma$,
one obtains a depth of the imaginary part in the order
of $W_0 \approx 100~{\rm{eV}}$ for the optical $s$- and $p$-wave potential.
This very weak imaginary part does not significantly influence the
calculated DC cross section.

A significant reduction of the DC cross section is only obtained when
$W_0$ is increased by more than three orders of magnitude. This means
that the calculations of the DC cross section in Fig.~\ref{fig:lambda}
are practically not influenced by the imaginary part of the potential.
Only potential resonances with very huge capture cross sections 
(with $\lambda \approx 1.08$) will be damped slightly by the imaginary 
part of the potential.

\section{Astrophysical reaction rates}
\label{sec:astro}
The astrophysical reaction rate $N_A < \sigma \cdot v >$
can easily be derived from the
Maxwellian averaged capture cross section 
$< \sigma > \, = \, < \sigma \cdot v > \, / \, v_{\rm T}$.
The Maxwellian averaged cross section 
of the reaction
$^{26}$Mg(n,$\gamma$)$^{27}$Mg was calculated from the following three 
ingredients. First, the thermal $s$-wave capture taken from Ref.~\cite{mugh}
was extrapolated to the thermonuclear energy region using the well-known
$1/v$ law. This $1/v$ behavior is reproduced by the DC calculations.
Therefore, the uncertainty of the $s$-wave contribution is below 5\%.
Second, the $p$-wave capture cross section which dominates at 
thermonuclear energies
is roughly proportional to the velocity $v$: 
\begin{equation}
\sigma = C \cdot \sqrt{E} .
\label{eq:pcontrib}
\end{equation}
The constant $C$ was determined
from the experimental data:
$C = 180 \pm 30~\rm{\mu b/\sqrt{MeV}}$.
Because only three experimental cross section data are not influenced
by resonances, and because one has to subtract the $s$-wave contribution
from the experimental capture cross section,
we estimate an uncertainty 
of less than 20\% for the $p$-wave contribution.
Note that the the $p$-wave contribution of the DC
for the data point at
$E_{\rm{n,lab}} = 208.3~{\rm{keV}}$ 
was also taken from Eq.~\ref{eq:pcontrib},
and the $s$- and $p$-wave contribution were subtracted
to determine the resonance strength of the
$E_{\rm{n,lab}} = 220~{\rm{keV}}$ resonance (see Sect.~\ref{subsec:results}).
Third, the resonance contributions were reduced by a factor of 2.8
compared to Ref.~\cite{weig76} (see discussion in Sect.~\ref{subsec:results}).
This factor was only determined for the $E_{\rm{n,lab}} = 220~{\rm{keV}}$
resonance, and the uncertainty of the {\bf sum} of the resonant
contributions is roughly a factor of 2.

\begin{figure}
\begin{center}
        \leavevmode
        \psfig{figure=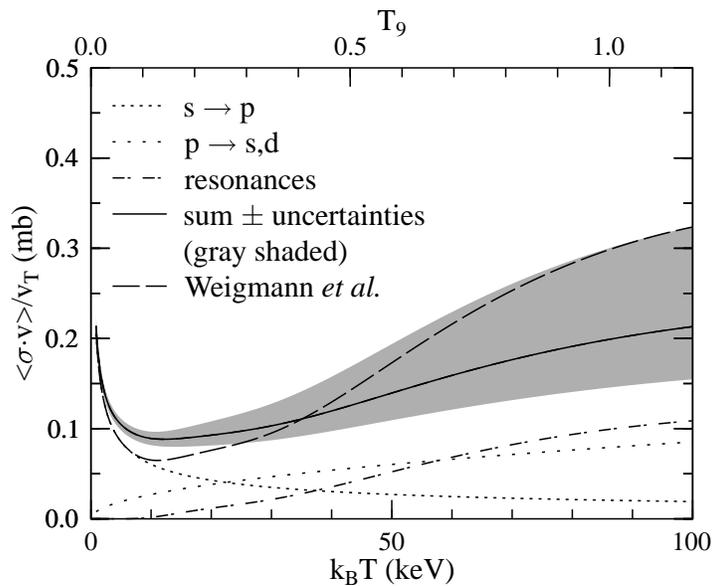,bbllx=30pt,bblly=120pt,bburx=600pt,bbury=310pt}
\end{center}
\caption{
	\label{fig:max}
	Maxwellian averaged capture cross section
	$< \sigma > \, = \, < \sigma \cdot v > \, / \, v_{\rm T}$,
	for the reaction
	$^{26}$Mg(n,$\gamma$)$^{27}$Mg. The total capture cross section is
	given by the sum of $s$-wave capture, $p$-wave capture, and resonant
	contributions. For comparison we also show the result from 
	Ref.~\protect\cite{weig76}.
}
\end{figure}

Finally, this leads to a Maxwellian averaged capture cross section
$< \sigma >$
as shown in Fig.~\ref{fig:max}. At low energies 
($k_B T < 10~{\rm{keV}}$)
$s$-wave capture is
dominating, and from about $k_B T \approx 20~{\rm{keV}}$ the $p$-wave
capture exceeds the $s$-wave contribution. The resonances become
dominant at higher energies ($k_B T \approx 50~{\rm{keV}}$).
By accident, our result is relatively close
to that of Ref.~\cite{weig76}: the sum of our measured $p$-wave capture
and the by a factor of 2.8 reduced resonant contributions
is somewhat larger at $k_B T < 30~{\rm{keV}}$ than
the calculation of Ref.~\cite{weig76} which neglected the $p$-wave
contribution, and at higher energies ($k_B T > 40~{\rm{keV}}$)
the result of Ref.~\cite{weig76} is higher than our new result
due to the overestimation of the resonance strengths in that work.

\section{Summary and conclusion}
\label{sec:summ}
There are three main results of this work: ($i$) the DC cross section
of the reaction $^{26}$Mg(n,$\gamma$)$^{27}$Mg was measured 
for the first time between the known resonances at
$E_{\rm{n,lab}} = 68.7~{\rm{keV}}$
and
$E_{\rm{n,lab}} = 220~{\rm{keV}}$
\cite{weig76}, 
($ii$) the properties of potential resonances
are discussed in detail for the $p$-wave capture of 
$^{26}$Mg(n,$\gamma$)$^{27}$Mg, and
($iii$) the resonance strength of the
$E_{\rm{n,lab}} = 220~{\rm{keV}}$ resonance
was determined to be $\omega \gamma = 1.34 \pm 0.24~{\rm{eV}}$
by a careful analysis of the neutron energy spectrum
$\Phi(E_{\rm{n,lab}})$.
This last result is in clear contradiction to the results 
of Ref.~\cite{weig76}.
Unfortunately, in Ref.~\cite{weig76} the experimental information on the
capture measurements is very limited, but it is stated in that work
that the enriched $^{26}$Mg- and $^{25}$Mg-samples were analyzed in the
same way. This fact makes the experimental results of Ref.~\cite{weig76}
for the capture reaction $^{25}$Mg(n,$\gamma$)$^{26}$Mg at least
questionable. A strongly reduced $^{25}$Mg(n,$\gamma$)$^{26}$Mg reaction
rate might have influence on the $s$-process nucleosynthesis because of the
role of $^{25}$Mg as neutron poison. In this paper we do not want to speculate
about astrophysical consequences of such a reduced reaction rate; instead,
we think that primarily a new measurement of the resonance properties 
of $^{25}$Mg is necessary to obtain reliable reaction rates.
Unfortunately, the reaction $^{25}$Mg(n,$\gamma$)$^{26}$Mg cannot 
be analyzed using our activation technique because the residual nucleus
$^{26}$Mg is stable.

\acknowledgements 
We would like to thank the technicians G.~Rupp, E.~Roller, E.-P.~Knaetsch,
and W.~Seith from the Van-de-Graaff laboratory at the Forschungszentrum
Karlsruhe for their support during the experiment and for the reliable beam.
This work was supported by Fonds zur F\"orderung der
Wissenschaftlichen Forschung in \"Osterreich (project S7307--AST),
Deutsche Forschungsgemeinschaft (DFG) (project Mo739/1),
and Volkswagen--Stiftung (Az: I/72286).

\begin{table}
\caption{\label{tab:tab1} 
Decay properties of the
residual nuclei $^{27}$Mg and $^{198}$Au.}
\begin{center}
\begin{tabular}{cccc}
Residual        & $T_{1/2}$                     & $E_\gamma$ 
        & Intensity per decay \\
nucleus         &                               & (keV)    
        & (\%) \\
\hline
$^{27}$Mg       & 9.458 $\pm$ 0.012 min & 843.76
        & 71.8 $\pm$ 0.4        \\
                &                               & 1014.44
        & 28.0 $\pm$ 0.4        \\
$^{198}$Au      & 2.69517 $\pm$ 0.00002 d       & 411.80
        & 95.50$\pm$0.096       \\
\end{tabular}
\end{center}
\end{table}

\begin{table}
\caption{\label{tab:tab2} 
Properties of the
magnesium samples and the gold foils.}
\begin{center}
\begin{tabular}{cccc}
Isotope         & Chemical              & Isotopic
        & mass \\
        & form                          & composition (\%)
        & (mg) \\
\hline
$^{26}$Mg       & MgO           & $98.79 \pm 1.90$ (enriched)
        & 29.20 $\pm$ 0.05 \\
$^{26}$Mg       & MgO           & $98.79 \pm 1.90$ (enriched)
        & 99.50 $\pm$ 0.20 \\
$^{26}$Mg       & metallic              & $11.01 \pm 0.02$ (natural)
        & 47.77 $\pm$ 0.02 \\
$^{197}$Au      &metallic               & 100 (natural)
        & 15.8 -- 16.5 \\
\end{tabular}
\end{center}
\end{table}

\begin{table}
\caption{
        \label{tab:tab3} 
        Experimental capture cross section $\sigma_{\rm{exp}}$
        of the reaction
        $^{26}$Mg(n,$\gamma$)$^{27}$Mg
        compared to the calculated DC cross section $\sigma_\gamma$
        calculated with $\lambda = 0.9694$ 
        (see Sect.~\protect\ref{sec:theo} and 
        Fig.~\protect\ref{fig:exp}).
}
\begin{tabular}{ccc}
$E_{\rm{n,lab}}$ (keV)  & $\sigma_{\rm{exp}}$(n,$\gamma$) ($\mu$b) 
          & $\sigma_\gamma$ ($\mu$b) \\
\hline
$kT = 25.3~{\rm{meV}}$  & $(39.0 \pm 0.8)$ mb \tablenotemark[1]
                                                & 32.5 mb  \\
$kT = 25~{\rm{keV}}$    & $124.2 \pm 7.5$       & 92.4     \\
33                      & $56.8 \pm 3.0$        & 98.1     \\
102                     & $82.1 \pm 4.5$        & 151.2    \\
151                     & $84.9 \pm 5.0$        & 189.9    \\
178                     & $188 \pm 28$          & 210.0    \\
208                     & $735 \pm 40$          & 236.4    \\
\end{tabular}
\tablenotetext[1]{from Ref.~\protect\cite{wal92}}
\end{table}

\begin{table}
\caption{
        \label{tab:spec}
        Spectroscopic factors of $^{27}$Mg = $^{26}$Mg $\otimes$ n
        taken from different experiments
        \protect\cite{meur74,turk88,pasch75,sinc76,koh80}
        and from shell-model calculations
        \protect\cite{meur74,moroz86,ben84}.
        Note that the spectroscopic factor of the state at $E_x =
        4149.8~{\rm{keV}}$ was determined only in one experiment.
        Additionally, the potential strength parameter $\lambda$ is given,
        which is adjusted to the binding energy of each bound state
        (see Sect.~\protect\ref{subsec:DC}).
}
\begin{tabular}{cccccc}
$E_x ({\rm{keV}})$      & $J^{\pi}$     
        & $C^2 S_{\rm{exp}}$    & $C^2 S_{\rm{th}}$
        & $C^2 S_{(n,\gamma)}$  & $\lambda$ \\
\hline
0       & $1/2^+$    & $0.44 - 1.07$ & $0.43 - 0.70$ & $-$ & 0.967          \\
984.7   & $3/2^+$    & $0.37 - 0.80$ & $0.28 - 0.45$ & $-$ & 0.947          \\
1698.0  & $5/2^+$    & $0.13 - 0.31$ & $0.02 - 0.14$ & $-$ & 0.922          \\
3559.5  & $3/2^-$    & $0.34 - 0.56$ & $-$           & $0.360 \pm 0.054$ 
    & 1.280 \\
3760.4  & $7/2^-$    & $0.40 - 0.80$ & $-$           & $-$ & 1.360          \\
4149.8  & $(5/2^+)$  & $0.03^*$      & $0.001 - 0.06$& $-$ & 0.827          \\
4827.3  & $(1/2^-)$  & $0.32-0.39^*$ & $-$           & $0.260 \pm 0.039^*$
    & 1.208 \\
\hline
\multicolumn{5}{l}{\small{$^*$ assuming $(J^\pi)$ as given in column 2}}
\end{tabular}
\end{table}

\begin{table}
\caption{
        \label{tab:res}
        Calculated resonance energies $E_{\rm{res}}$ and widths $\Gamma$
        of a $p$-wave potential resonance for 
        potential strength parameters $\lambda$ from 0.90 to 1.10. For
        $\lambda < 0.9$ the potential resonance becomes very broad.
}
\begin{tabular}{cccc}
$\lambda$       
        & $E_{\rm{res}}~{\rm{(keV)}}$   
        & $\Gamma~{\rm{(keV)}}$
        & $\delta_{l=1}(E_{\rm{res}})$ (deg)\\
\hline
0.90            & 795                   & 4130          & 16$^o$\\
0.92            & 755                   & 3170          & 20$^o$\\
0.94            & 705                   & 2400          & 24$^o$\\
0.96            & 647                   & 1785          & 28$^o$\\
0.98            & 562                   & 1284          & 32$^o$\\
1.00            & 485                   & 878           & 38$^o$\\
1.02            & 381                   & 553           & 44$^o$\\
1.04            & 274                   & 300           & 51$^o$\\
1.06            & 156.5                 & 116           & 62$^o$\\
1.08            & 28.6                  & 8.2           & 78$^o$\\
1.10            & -135.3\tablenotemark[1]               
                                        & --            & --    \\
\end{tabular}
\tablenotetext[1]{bound state with $Q=2N+L=3$, $N=1$, $L=1$}
\end{table}

\end{document}